\documentclass{cargese}
\usepackage{graphicx}
\usepackage{amsmath}
\usepackage{amssymb}
\let\footnote\savefootnote\let\footnotetext\savefootnotetext \let\footnoterule\savefootnoterule \normallatexbib

\begin{document}

\articletitle[Braneworld cosmology]{Braneworld cosmology \\ almost without branes\footnote{Talk presented at Carg\`{e}se Summer School 2004, NATO Advanced Study Institute, Carg\`{e}se, France, June 7-19 2004, {\it String Theory: From Gauge Interactions to Cosmology}.}}

%\chaptitlerunninghead{}

\author{Gianluca Calcagni}
\affil{Dipartimento di Fisica, Universit\`a di Parma}   
\affil{INFN -- Gruppo collegato di Parma \\ Parco Area delle Scienze 7/A, I-43100 Parma, Italy}
\email{calcagni@fis.unipr.it}

\begin{abstract}
We review some general aspects of braneworld cosmologies in which an inflationary period driven by a scalar field confined on the brane is described by a nonstandard effective Friedmann equation. The perturbation spectra, consistency equations and observational consequences of these models are considered.
\end{abstract}

%%%%%%%%%%%%%%%%%%%%%%%%%%%%%%%%%%%%%%%%%%%%%%%%%%%%%%%%%%%%%%%%%%%%%%%%%%%%%%%%%%%%%%%%%%%%%%%%%%%%%%%%%%%%%%%%%%%%%%%%%%%%%%%%%%%%%%%%%%%%%%%%%%%%%%%%%%%%%%%%%%%%%%%%%%%%%%%%%%%%%%%%%%%%%%%%%%%%%%%%%%%%%%%%%%%%%%%%%%%%%%%%%%%%%%%%%%%%%%%%

\section{Introduction}

Motivated by recent developments in string, superstring and M theory, several models for a multidimensional target spacetime have been proposed. Among them, particular attention has been devoted to brane-world scenarios, according to which the visible universe is a (3+1)-dimensional variety (a 3-brane) embedded in a bulk with some either non-compact or compactified extra dimensions. Typically, the background metric on the brane is assumed to be the Friedmann-Robertson-Walker (FRW) metric and the Einstein equations are modified in accordance with the gravity model describing spacetime. Their projection on the brane results in the basic FRW equations for the cosmological evolution.  
For an introduction to the subject and some lists of references, see \cite{bw}.

In this paper we review how to look for cosmic signatures of high-energy, higher-derivative gravity models. In particular, the construction of a nontrivial set of consistency equations permits us to compare theoretical predictions with the perturbation spectra of the cosmic microwave background (CMB). It turns out that CMB experiments of this and next generation might be able to discriminate between the standard four-dimensional lore and the braneworld paradigm. In Secs. \ref{setup} and \ref{roll} we introduce the basic ingredients of the patch and slow-roll formalisms, taking as examples the five-dimensional Randall-Sundrum (RS) scenarios and their Gauss-Bonnet (GB) generalization. In Sec. \ref{pert} we outline some results on cosmological perturbations in the presence of an extra dimension and find their observational consequences. Conclusions are in Sec. \ref{concl}

%%%%%%%%%%%%%%%%%%%%%%%%%%%%%%%%%%%%%%%%%%%%%%%%%%%%%%%%%%%%%%%%%%%%%%%%%%%%%%%%%%%%%%%%%%%%%%%%%%%%%%%%%%%%%%%%%%%%%%%%%%%%%%%%%%%%%%%%%%%%%%%%%%%%%%%%%%%%%%%%%%%%%%%%%%%%%%%%%%%%%%%%%%%%%%%%%%%%%%%%%%%%%%%%%%%%%%%%%%%%%%%%%%%%%%%%%%%%%%%%

\section{Setup} \label{setup}

One of the first problems one has to deal with when constructing braneworld models is how to stabilize the extra dimension. This can be achieved in a number of ways; in the RS example, Goldberger and Wise have provided a mechanism according to which a 5D massive scalar is put into the bulk with a potential of the same order of the brane tension $\lambda$ \cite{GW}. If the energy density $\rho$ on the brane is smaller than the characteristic energy of the scalar potential, $\rho/V \sim \rho/\lambda \ll 1$, then the radion is stabilized and one gets the standard Friedmann equation $H^2 \propto \rho$ on the brane. On the contrary, if the brane energy density is comparable with the stabilization potential, $\rho/\lambda \gtrsim 1$, the bulk backreacts because it feels the presence of the brane matter, the minimum of the potential is shifted and the well-known quadratic corrections to the Friedmann equation arise:
\begin{equation}
H^2 = \frac{\kappa_4^2}{6 \lambda} \rho (2 \lambda+\rho)\,,	
\end{equation}
where $H$ is the effective Hubble rate experienced by an observer on the brane and $\kappa_4$ is the 4D gravitational coupling. 

The RS model can be viewed as a particular energy limit of a Gauss-Bonnet braneworld, characterized by the 5D Planck mass $M_5\equiv\kappa_5^{-2/3}$ and a Friedmann equation
\begin{equation} 
H^2=\frac{c_+ + c_- -2}{8\alpha}\,,
\end{equation}
where $\alpha=1/(8g_s^2)$ is the Gauss-Bonnet coupling ($g_s$ is the string coupling) and, defining $\sqrt{\alpha/2}\,\kappa_5^2 \equiv \sigma_0^{-1}$,
\begin{equation}
c_\pm = \left\{\left[\left(1+4\alpha\Lambda_5/3\right)^{3/2}+\left(\sigma/\sigma_0\right)^2\right]^{1/2} \pm \sigma/\sigma_0\right\}^{2/3}.
\end{equation}
$\Lambda_5<0$ is the bulk cosmological constant and $\sigma$ is the matter energy density which is decomposed into a matter contribution plus the brane tension $\lambda$: $\sigma=\rho+\lambda$. When $\sigma/\sigma_0 \gg 1$, one gets the ``pure Gauss-Bonnet'' high-energy regime,
\begin{equation}
H^2= \left(\frac{\kappa_5^2}{16\alpha}\right)^{2/3}\rho^{2/3}\,.
\end{equation}
When the energy density is far below the 5D scale but $\rho \gg \lambda$, one recovers the Friedmann equation of the RS scenario with vanishing 4D cosmological constant, provided that some relations among the parameters of the action are satisfied. 

Here we shall consider nonstandard cosmological evolutions on the brane and extend the RS and GB discussion to arbitrary scenarios we dubbed ``patch cosmologies'' \cite{cal3}, with
\begin{equation} \label{FRW}
H^2=\beta_q^2 \rho^q\,.
\end{equation}
$\beta_q$ is a constant and the exponent $q$ is equal to 1 in the pure 4D (radion-stabilized) regime, $q=2$ in the high-energy limit of the RS braneworld and $q=2/3$ in the high-energy limit of the GB scenario. In order to simplify the framework, we make the following assumptions:
\begin{enumerate}
\item There is a confinement mechanism such that matter lives on the brane only, while gravitons are free to propagate in the bulk. This is guaranteed as long as $\rho<M_5^4$.
\item The contribution of the Weyl tensor is neglected.
\item The contribution of the anisotropic stress is neglected.
\item We concentrate on the large-scale limit of the cosmological perturbations.
\end{enumerate}
Assumption 2 closes the system of equations on the brane and sets aside the nonlocal contributions from the bulk, while assumptions 3 and 4 reduce the number of degrees of freedom of gauge invariant scalar perturbations. 

This list might seem too restrictive and to spoil almost all the interesting features of the model. However, assumptions 4 and 2 nicely fit in the inflationary regime, since the long wavelength region of the spectrum, corresponding to the Sachs-Wolfe plateau, encodes the main physics of the inflationary era. Moreover, the dark radiation term, which is the simplest contribution of the Weyl tensor, scales as $a^{-4}$ and is exponentially damped during the accelerated expansion. Finally, bulk physics mainly affects the small-scale/late-time cosmological structure and can be consistently neglected during inflation. This is a highly nontrivial result which has been confirmed with several methods both analytically and numerically \cite{bu}.

Imposing a perfect fluid on the brane with equation of state $p=w\rho$, the continuity equation governing the cosmological dynamics is the same as in four dimensions, thanks to assumption 2:
\begin{equation}
\dot{\rho}+3H (\rho+p)=0\,.
\end{equation}
There are two candidates for the role of inflation. The first one is an ordinary scalar field $\phi$ with energy density and pressure
\begin{equation}
\rho=\dot{\phi}^2/2+V(\phi)=p+2V(\phi)\,.
\end{equation}
The second one is a Dirac-Born-Infeld (DBI) tachyon $T$ such that
\begin{eqnarray} \label{tac}
\rho &=& V(T)/c_S=-V(T)^2/p\,,\\
c_S &\equiv& \sqrt{1-\dot{T}^2}\,.
\end{eqnarray}
From a string-theoretical point of view, the evolution of the tachyon proceeds up to the condensation $\dot{T}\to 1$ into the closed string vacuum, where no signal of open excitations propagates (Carrollian limit). Also, near the minimum the strong coupling regime emerges, $g_s=O(1)$, and the perturbative description implicit in the DBI action may fail down. However, from a cosmological perspective Eq. (\ref{tac}) is a toy model and, like in standard inflation, an additional reheating mechanism around the condensation is required for gracefully exiting the inflationary period. Here we will not consider this and other (indeed solvable) problems concerning the tachyon and just implement the DBI action in the cosmological dynamics as an alternative model of inflation.

%%%%%%%%%%%%%%%%%%%%%%%%%%%%%%%%%%%%%%%%%%%%%%%%%%%%%%%%%%%%%%%%%%%%%%%%%%%%%%%%%%%%%%%%%%%%%%%%%%%%%%%%%%%%%%%%%%%%%%%%%%%%%%%%%%%%%%%%%%%%%%%%%%%%%%%%%%%%%%%%%%%%%%%%%%%%%%%%%%%%%%%%%%%%%%%%%%%%%%%%%%%%%%%%%%%%%%%%%%%%%%%%%%%%%%%%%%%%%%%%

\section{Slow-roll parameters} \label{roll}

Let $\psi$ denote the inflaton field irrespectively of its action. Expressions involving $\psi$ will be valid for both the ordinary scalar and the tachyon. The first-order slow-roll (SR) parameters are defined as
\begin{equation} \label{1SR}
\epsilon \equiv -\frac{\dot{H}}{H^2}\,,\qquad \eta \equiv -\frac{\ddot{\psi}}{H\dot{\psi}}\,,
\end{equation}
together with their evolution equations with respect to synchronous time
\begin{equation}
\dot{\epsilon} = H\epsilon \left[\left(2-\widetilde{\theta}\right)\,\epsilon-2\eta\right]\,,\qquad
\dot{\eta}     =  H\left(\epsilon\eta-\xi^2\right)\,,
\end{equation}
where $\xi^2 \equiv (\ddot{\psi}/\dot{\psi})^\cdot/H^2$ is a second-order parameter, in the sense that it appears only in expressions which are $O(\epsilon^2,\eta^2,\epsilon\eta)$. Here $\widetilde{\theta}=2$ for the tachyon field and $\widetilde{\theta}=\theta \equiv 2(1-q^{-1})$ for the ordinary scalar field (4D: $\theta=0$; RS: $\theta=1$; GB: $\theta=-1$). Note that each time derivative of the SR parameters increases the order of the SR expressions by one.

One can construct infinite towers of SR parameters encoding the full dynamics of the inflationary model. For instance, Eq. (\ref{1SR}) provides the first entries of the ``Hubble" SR tower; another, sometimes more convenient tower is the potential tower defined as
\begin{eqnarray}
\epsilon_{\text{\tiny $\phi V$},0} &\equiv& \frac{q}{6\beta_q^2}\frac{V'^2}{V^{1+q}}\,,\\
\epsilon_{\text{\tiny $\phi V$},n} &\equiv& \frac{1}{3\beta_q^2}\left[\frac{V^{(n+1)}(V')^{n-1}}{V^{nq}}\right]^{1/n}\,, \qquad n \geq 1\,,
\end{eqnarray}
in the case of the normal scalar field. The potential tower can be related to the Hubble tower by approximated relations \cite{cal3}.

The first SR parameter is actually the time derivative of the Hubble radius $R_H\equiv H^{-1}$. Because of its purely geometrical content, it cannot be implemented in these SR towers recursively. By definition, there is inflation when $\epsilon < 1$:
\begin{equation} \label{infl}
\frac{\ddot{a}}{a} = H^2 (1-\epsilon)\,.
\end{equation}
Under the \emph{slow-roll approximation}, if the potential term dominates over the kinetic term, then the inflaton slowly rolls down its potential, $\epsilon,\eta \ll 1$, and the perfect fluid mimics that of a cosmological constant, $p \approx -\rho$. Deviations from the de Sitter behaviour generate large-scale perturbations which explain the anisotropies in the CMB.

%%%%%%%%%%%%%%%%%%%%%%%%%%%%%%%%%%%%%%%%%%%%%%%%%%%%%%%%%%%%%%%%%%%%%%%%%%%%%%%%%%%%%%%%%%%%%%%%%%%%%%%%%%%%%%%%%%%%%%%%%%%%%%%%%%%%%%%%%%%%%%%%%%%%%%%%%%%%%%%%%%%%%%%%%%%%%%%%%%%%%%%%%%%%%%%%%%%%%%%%%%%%%%%%%%%%%%%%%%%%%%%%%%%%%%%%%%%%%%%%

\section{Cosmological perturbations: theory and observations} \label{pert}

Quantum fluctuations of the scalar field governing the accelerated era are inflated to cosmological scales because of the superluminal expansion. They constitute the seeds of both the small anisotropies observed in the microwave sky and the large-scale nonlinear structures around which gravitating matter organizes itself. For an introduction of the subject in the general relativistic case, see \cite{MFB}. The standard procedure to adopt in order to compute the perturbation spetrum is: (a) Write the linearly perturbed metric in terms of gauge-invariant scalar quantities. (b) Compute the effective action of the scalar field fluctuation and the associated equation of motion. (c) Write the perturbation amplitude as a function of an exact solution of the equation of motion with constant SR parameters. (d) Perturb this solution with small variations of the parameters.

In scenarios with an extra dimension the full computation is very nontrivial due to either the extra degrees of freedom in the 5D metric and the complicated geometrical background on which to solve the Einstein equations coupled with the junction conditions on the brane. However, as explained above things become simpler when going to the large-scale limit. In this case, several arguments show that the resulting spectra are, to lowest SR order,
\begin{eqnarray}
A       &=& \frac{k}{5\pi z}\,,\\
z(\phi) &=& \frac{a\dot{\phi}}{H}\,,\\
z(T)    &=& \frac{a\dot{T}}{c_S\beta_q^{1/q} H^{\theta/2}}\,,\\
z(h)    &=& \frac{\sqrt{2}a}{\kappa_4 F_q}\,,\\
F^2_q &\equiv& \frac{3q\beta_q^{2-\theta}H^\theta}{\zeta_q\kappa_4^2}\,,
\end{eqnarray}
where $A(h)=A_t$ is the tensor spectrum of the gravitational sector and $\zeta_q$ is a numerical constant which depends on the concrete gravity model one is considering: it is $\zeta_1=1=\zeta_{2/3}$ for the 4D and GB cases and $\zeta_2=2/3$ for RS \cite{zeta}.

To lowest order, the scalar and tensor spectral indices are first order in the SR parameter, $n_t \equiv d \ln A_t^2/d \ln k \sim O(\epsilon) \sim n_s-1 \equiv d\ln A_s^2/d \ln k$, while their running $\alpha_{s,t} \equiv d n_{s,t}/d \ln k$ is second order. Here $k$ is the comoving wave number of the perturbation and the subscripts $s$ and $t$ refer to scalar and tensor perturbations, respectively. In the case of exact scale invariance, $n_s=1$ and $n_t=0$. The tensor-to-scalar ratio is
\begin{equation}
r \equiv A_t^2/A_s^2=\epsilon/\zeta_q+O(\epsilon^2)\,.
\end{equation}
Combining the SR expressions of the observables, one gets the consistency equations
\begin{eqnarray}
n_t            &=& -(2+\theta)\zeta_q r+O(\epsilon^2)\,, \label{1ce}\\
\alpha_t &=& (2+\theta)\zeta_qr[(2+\theta)\zeta_qr+(n_s-1)]\,,\\
\alpha_s(\phi) &\approx& \zeta_q r [4(3+\theta)\zeta_qr+5(n_s-1)]\,,\\
\alpha_s(T)    &\approx& (3+\theta)\zeta_q r [(2+\theta)\zeta_qr+(n_s-1)]\,.
\end{eqnarray}
The key point is that the set of consistency relations is not degenerate when considering different patches $\theta$ and $\theta'$. The only known (accidental) degeneracy is for Eq. (\ref{1ce}) in the RS and 4D case, where $n_t=-2r$ at first SR order. However, the second-order version of this equation together with the expressions for the runnings definitely break the degeneracy. This implies that, at least in principle, braneworld scenarios can be discriminated between each other.

To quantify the effect of the extra dimension, we can use the recent CMB data coming from WMAP \cite{wmap}. With the upper bound $r < 0.06$ for the tensor-to-scalar ratio and the best-fit value $n_s \approx 0.95$ for the scalar spectral index, the relative scalar running in two different patches is
\begin{equation}
\alpha_s^{(\theta,\psi)}-\alpha_s^{(\theta',\psi')} \sim O(10^{-2})\,, 
\end{equation}
which is close to the WMAP estimate of the experimental error. This estimate will be highly improved by either the updated WMAP data set and near-future experiments, including the European Planck satellite, for which the forecast precision should be ameliorated by one order of magnitude, $\Delta\alpha_s \sim O(10^{-3})$. 

%%%%%%%%%%%%%%%%%%%%%%%%%%%%%%%%%%%%%%%%%%%%%%%%%%%%%%%%%%%%%%%%%%%%%%%%%%%%%%%%%%%%%%%%%%%%%%%%%%%%%%%%%%%%%%%%%%%%%%%%%%%%%%%%%%%%%%%%%%%%%%%%%%%%%%%%%%%%%%%%%%%%%%%%%%%%%%%%%%%%%%%%%%%%%%%%%%%%%%%%%%%%%%%%%%%%%%%%%%%%%%%%%%%%%%%%%%%%%%%%

\section{Conclusions} \label{concl}

In this paper we have summarized some results on braneworld inflation and their observable consequences. We have not presented a full 5D calculation but we expect that bulk physics would not dramatically improve large-scale results \cite{bu}. The study of the microwave background could give the first clues of a wider spacetime in the next years or even months.
 
In addition to the brane conjecture, one may insert other exotic ingredients, borrowed from string and M theory, that may give rise to characteristic predictions, although at the price of increasing the number and complexity of concurring models. For instance, the introduction of a noncommutative scale can generate a blue-tilted spectrum and explain, at least partially, the low-multipole suppression of the CMB spectrum detected by WMAP (e.g. \cite{nonc}).

It would be interesting to find new cosmological scenarios with $\theta\neq 0,\pm 1$ and exploit the compact formalism provided by the patch formulation of the cosmological dynamics. Certainly there could be a lot of work for M/string theorists in this direction.

A final important question is in order: Will the CMB be the smoking gun of extra dimensions? In the context of the patch formalism the answer, unfortunately, is no. Some general relativistic models may predict a set of values for the observables $\{n_t,n_s,r,\alpha_s,\dots\}$  close to that of a braneworld within the experimental sensitivity. Even noncommutativity may not escape this ``cosmic degeneracy'' since, for example, a blue-tilted spectrum can be achieved by the 4D hybrid inflation. So we can talk about clues but not proofs about high-energy cosmologies when examining the experimental data. The subject has to be further explored in a more precise way than that provided here in order to find out more compelling and sophisticated predictions, extending the discussion also to the small-scale region of the spectrum.

%%%%%%%%%%%%%%%%%%%%%%%%%%%%%%%%%%%%%%%%%%%%%%%%%%%%%%%%%%%%%%%%%%%%%%%%%%%%%%%%%%%%%%%%%%%%%%%%%%%%%%%%%%%%%%%%%%%%%%%%%%%%%%%%%%%%%%%%%%%%%%%%%%%%%%%%%%%%%%%%%%%%%%%%%%%%%%%%%%%%%%%%%%%%%%%%%%%%%%%%%%%%%%%%%%%%%%%%%%%%%%%%%%%%%%%%%%%%%%%%

\begin{acknowledgments}
It is a pleasure to thank the organizers for their kind hospitality at Carg\`{e}se Summer School 2004.
\end{acknowledgments}

%%%%%%%%%%%%%%%%%%%%%%%%%%%%%%%%%%%%%%%%%%%%%%%%%%%%%%%%%%%%%%%%%%%%%%%%%%%%%%%%%%%%%%%%%%%%%%%%%%%%%%%%%%%%%%%%%%%%%%%%%%%%%%%%%%%%%%%%%%%%%%%%%%%%%%%%%%%%%%%%%%%%%%%%%%%%%%%%%%%%%%%%%%%%%%%%%%%%%%%%%%%%%%%%%%%%%%%%%%%%%%%%%%%%%%%%%%%%%%%%


\begin{chapthebibliography}{50}

\bibitem{bw}   V.A.  Rubakov, Phys. Usp. \textbf{44} (2001) 871, hep-ph/0104152; R. Maartens, Living Rev. Relativity \textbf{7} (2004) 1, gr-qc/0312059; P. Brax, C. van de Bruck, and A.-C. Davis, hep-th/0404011.

\bibitem{GW}	 W.D. Goldberger and M.B. Wise, Phys. Rev. Lett. \textbf{83} (1999) 4922, hep-ph/9907447; W.D. Goldberger and M.B. Wise, Phys. Lett. B \textbf{475} (2000) 275, hep-ph/9911457.
							
\bibitem{cal3} G. Calcagni, Phys. Rev. D \textbf{69} (2004) 103508, hep-ph/0402126.							 
													 
\bibitem{bu}   D.S. Gorbunov, V.A. Rubakov, and S.M. Sibiryakov, J. High Energy Phys. \textbf{10} (2001) 015, hep-th/0108017; C. Gordon and R. Maartens, Phys. Rev. D \textbf{63} (2001) 044022, hep-th/0009010; K. Ichiki, M. Yahiro, T. Kajino, M. Orito, and G.J. Mathews, Phys. Rev. D \textbf{66} (2002) 043521, astro-ph/0203272; B. Leong, A. Challinor, R. Maartens, and A. Lasenby, Phys. Rev. D \textbf{66} (2002) 104010, astro-ph/0208015; K. Koyama, Phys. Rev. Lett. \textbf{91} (2003) 221301, astro-ph/0303108; K. Koyama, D. Langlois, R. Maartens, and	D. Wands, hep-th/0408222.	
 							 
\bibitem{MFB}  V.F. Mukhanov, H.A. Feldman, and R.H. Brandenberger, Phys. Rep. \textbf{215} (1992) 203. 	

\bibitem{zeta} D. Langlois, R. Maartens, D. Wands, Phys. Lett. B \textbf{489} (2000) 259, hep-th/0006007; J.F. Dufaux, J.E. Lidsey, R. Maartens, M. Sami, hep-th/0404161; G. Calcagni, hep-ph/0406057.

\bibitem{wmap} C.L. Bennett \textit{et al.}, Astrophys. J., Suppl. Ser. \textbf{148} (2003) 1, astro-ph/0302207; 							 				 D.N. Spergel \textit{et al.}, Astrophys. J., Suppl. Ser. \textbf{148} (2003) 175, astro-ph/0302209;							 			 H.V. Peiris \textit{et al.}, Astrophys. J., Suppl. Ser. \textbf{148} (2003) 213, astro-ph/0302225;
							 S.L. Bridle, A.M. Lewis, J. Weller, and G. Efstathiou, Mon. Not. R. Astron. Soc. \textbf{342} (2003) L72, astro-ph/0302306. 					 
							 
\bibitem{nonc} G. Calcagni, Phys. Rev. D \textbf{70} (2004) 103525, hep-th/0406006; G. Calcagni and S. Tsujikawa, Phys. Rev. D \textbf{70} (2004) 103514, astro-ph/0407543.
							 
\end{chapthebibliography}
\end{document}